\documentclass[11pt]{article}
\usepackage[margin=3.4cm]{geometry}
\usepackage{times}
\usepackage{amsmath,amssymb}
\usepackage{setspace,subfig}
\usepackage{natbib}
\usepackage{authblk}
\usepackage{dcolumn}
\usepackage{longtable}
\usepackage{relsize}
\usepackage{booktabs}
\usepackage{pgf}
\usepackage{tikz}
\usetikzlibrary{arrows,graphs,shapes.arrows,shapes.geometric,shapes.misc,
shapes.multipart,backgrounds,decorations.pathmorphing,positioning,fit,patterns}
\usepackage{tkz-graph}
\usepackage[nomessages]{fp} \usepackage{multirow}
\usepackage{setspace}
\usepackage{times}
\usepackage{mathtools,bbm}

\newcolumntype{d}[1]{D{.}{.}{#1}}
\renewcommand{\harvardurl}[1]{\textbf{URL:} \url{#1}}

\DeclareMathOperator{\pr}{\rm{pr}}
\DeclareMathOperator{\E}{\text{E}}

\tikzset{node distance=5mm,
 terminal/.style={
rounded rectangle,
minimum size=3mm,
very thick,draw=black!50,
top color=white,bottom color=black!20, font=\ttfamily\scriptsize},
 terminal2/.style={
rounded rectangle,
minimum size=2.5mm,
inner sep=1pt,
thick,draw=black!50,
top color=white,bottom color=black!20, font=\ttfamily\tiny},
}

\makeatletter
\begin{document}

\title{Discussion of `Estimating time-varying causal excursion effect in
mobile health with binary outcomes'\\
by {T.~Qian, et al.}}
\author[1]{F. Richard Guo}
\author[1]{Thomas S. Richardson}
\author[2]{James M. Robins}
\affil[1]{Department of Statistics, University of Washington, Box 354322, Washington 98195, U.S.A.
		\texttt{\{ricguo, thomasr\}@uw.edu}}

\affil[2]{Department of Epidemiology, Harvard T.H. Chan School of Public Health,
		677 Huntington Avenue, Boston, Massachusetts 02115, U.S.A.
		\texttt{robins@hsph.harvard.edu}}
\date{}
\maketitle

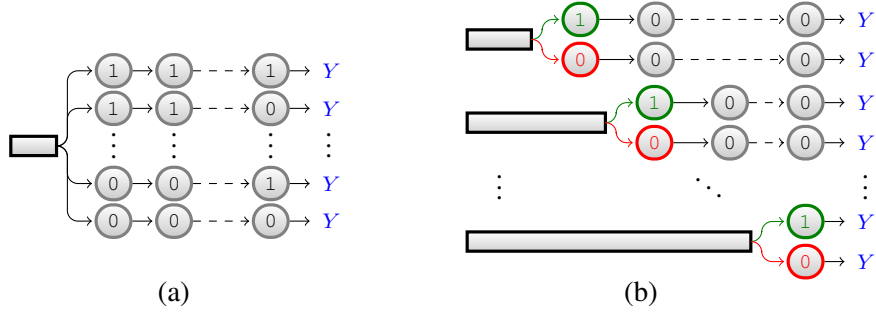
\begin{figure}
\centering
\begin{tikzpicture}[every new ->/.style={shorten >=1pt},
graphs/every graph/.style={edges=rounded corners},
vh path/.style={to path={-- ++(3.5pt,0) |- (\tikztotarget)}}
]
\node(cy){};
\node (ui1) [rectangle, top color=white, bottom color=black!20, draw, very thick, draw=black, left= of cy, xshift=-60pt] {\kern 10 pt};
\node (0y) [font=\scriptsize, blue, right = of cy,yshift=-10 mm] {$Y$}; 
\node (0b) [terminal, left = of 0y,xshift=2 mm] {0}; 
\node (0a) [terminal, left = of 0b,xshift=-8 pt] {0}; 
\node (0) [terminal, left = of 0a, xshift=2 mm] {0}; 
\node (labela) [below = of 0a,yshift=2pt] {(a)};     \graph{
 (ui1.east) ->[vh path]       (0);
  (0) -> (0a);
  (0a) ->[dashed]  (0b);
   (0b) -> (0y);
};
\node (2y) [font=\scriptsize, blue, right = of cy,yshift=5 mm] {$Y$}; 
\node (2b) [terminal, left = of 2y,xshift=2 mm] {0}; 
\node (2a) [terminal, left = of 2b,xshift=-8 pt] {1}; 
\node (2) [terminal, left = of 2a, xshift=2 mm] {1}; 
\graph{
    (ui1.east) ->[vh path]       (2);
      (2) -> (2a);
  (2a) ->[dashed]  (2b);
   (2b) -> (2y);
   };
\node (1y) [font=\scriptsize, blue, right = of cy,yshift=10 mm] {$Y$}; 
\node (1b) [terminal, left = of 1y,xshift=2 mm] {1}; 
\node (1a) [terminal, left = of 1b,xshift=-8 pt] {1}; 
\node (1) [terminal, left = of 1a, xshift=2 mm] {1}; 
\graph{
    (ui1.east) ->[vh path]       (1);
      (1) -> (1a);
  (1a) ->[dashed]  (1b);
   (1b) -> (1y);
   };
\node (3y) [font=\scriptsize, blue, right = of cy,yshift=-5 mm] {$Y$}; 
\node (3b) [terminal, left = of 3y,xshift=2 mm] {1}; 
\node (3a) [terminal, left = of 3b,xshift=-8 pt] {0}; 
\node (3) [terminal, left = of 3a, xshift=2 mm] {0}; 
\graph{
    (ui1.east) ->[vh path]       (3);
      (3) -> (3a);
  (3a) ->[dashed]  (3b);
   (3b) -> (3y);
   };
\node (dy) [below = of 1y, yshift=7pt,xshift=-1pt] {$\vdots$}; 
\node (db) [below = of 1b, yshift=7pt] {$\vdots$}; 
\node (da) [below = of 1a, yshift=7pt] {$\vdots$}; 
\node (d) [below = of 1, yshift=7pt] {$\vdots$}; 
\begin{scope}[xshift=8cm,yshift=4cm]
\node (labelb) [right = of labela,xshift=5cm] {(b)};     \foreach \x in {0,7}
{
\FPeval{Width}{(45-4*\x)}; \begin{scope}[yshift=-105+4.5*\x pt]
\node(cy){};
\node (0y) [font=\scriptsize, blue, above = of cy,yshift=-6mm] {$Y$}; 
\node (0b) [terminal, left = of 0y,xshift=2 mm] {0}; 
\node (0a) [terminal, left = of 0b,xshift=-4*\x pt] {0}; 
\node (0) [terminal, green!50!black, left = of 0a, xshift=0pt] {1}; 
\node (1y) [font=\scriptsize, blue, below = of cy,yshift=6mm] {$Y$}; 
\node (1b) [terminal, left = of 1y,xshift=2 mm] {0}; 
\node (1a) [terminal, left = of 1b,xshift=-4*\x pt] {0}; 
\node (1) [terminal, red, left = of 1a,xshift=0pt] {0}; 
\node (ui1) [rectangle, top color=white, bottom color=black!20, draw, very thick, draw=black, left= of cy, xshift=-80-4*\x pt] {\kern \Width pt};
\graph{
 (ui1.east) ->[vh path,green!50!black]       (0);
  (ui1.east) ->[vh path,red]       (1);
   (0) ->  (0a);
  (0a) ->[dashed]  (0b);
   (0b) -> (0y);
  (1) ->  (1a);
  (1a) ->[dashed]  (1b);
   (1b) -> (1y);
  };
\end{scope} 
}
\begin{scope}\node (d) [below = of ui1,yshift=-23pt] {$\vdots$};
\node (d2) [right = of d,xshift=50pt] {$\ddots$};
\node (d3) [right = of d2,xshift=30pt] {$\vdots$};
\end{scope}

\begin{scope}[yshift=-105-45 pt]
\node(cy){};
\node (0y) [font=\scriptsize, blue, above = of cy,yshift=-6mm] {$Y$}; 
\node (0) [terminal, green!50!black, left = of 0y, xshift=2 mm] {1}; 
\node (1y) [font=\scriptsize, blue, below = of cy,yshift=6mm] {$Y$}; 
\node (1) [terminal, red, left = of 1y,xshift=2mm] {0}; 
\node (ui1) [rectangle, top color=white, bottom color=black!20, draw, very thick, draw=black, left= of cy, xshift=-25 pt] {\kern 100 pt};
\graph{
 (ui1.east) ->[vh path,green!50!black]       (0);
  (ui1.east) ->[vh path,red]       (1);
(0) -> (0y);
(1) -> (1y);
  };
\end{scope} 
\end{scope}
\end{tikzpicture}
\caption{A {\em marginal structural model} (a) specifies the expected counterfactual outcome for $Y=Y_{T}$ under every static regime $(a_1,\ldots, a_T)$ given baseline covariates indicated by the grey rectangle; 
A {\em structural nested mean model} (b) specifies contrasts for all times $t$ giving the difference in expected counterfactual outcome from receiving a final blip of treatment at time $t$, given all treatment and covariates prior to $t$.\label{fig:msm-snmm}}
\end{figure}

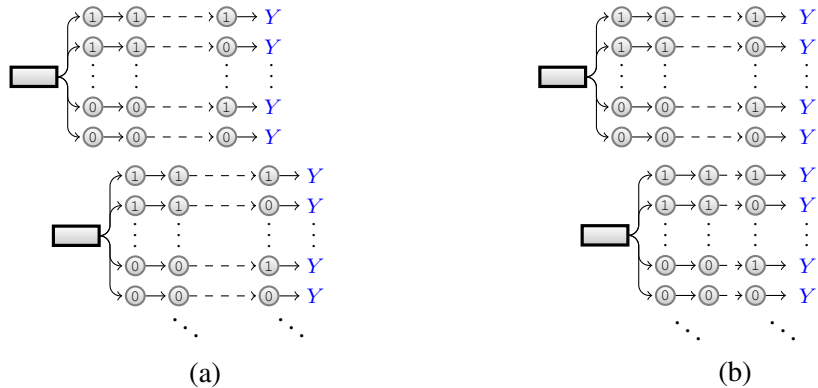
\begin{figure}
\centering
\begin{tikzpicture}[every new ->/.style={shorten >=1pt},
graphs/every graph/.style={edges=rounded corners},
vh path/.style={to path={-- ++(3.5pt,0) |- (\tikztotarget)}},
vh path2/.style={to path={|- (\tikztotarget)}}
]
\begin{scope}[yshift=-5pt,xshift=200pt]
\node(cy){};
\node (ui1) [rectangle, top color=white, bottom color=black!20, draw, very thick, draw=black, left= of cy, xshift=-40pt] {\kern 10 pt};
\node (0y) [font=\scriptsize, blue, right = of cy,yshift=-8 mm] {$Y$}; 
\node (0b) [terminal2, left = of 0y,xshift=2 mm] {0}; 
\node (0a) [terminal2, left = of 0b,xshift=-12 pt] {0}; 
\node (0) [terminal2, left = of 0a, xshift=2 mm] {0}; 
\graph{
 (ui1.east) ->[vh path]       (0);
  (0) -> (0a);
  (0a) ->[dashed]  (0b);
   (0b) -> (0y);
};
\node (2y) [font=\scriptsize, blue, right = of cy,yshift=4 mm] {$Y$}; 
\node (2b) [terminal2, left = of 2y,xshift=2 mm] {0}; 
\node (2a) [terminal2, left = of 2b,xshift=-12 pt] {1}; 
\node (2) [terminal2, left = of 2a, xshift=2 mm] {1}; 
\graph{
    (ui1.east) ->[vh path]       (2);
      (2) -> (2a);
  (2a) ->[dashed]  (2b);
   (2b) -> (2y);
   };
\node (1y) [font=\scriptsize, blue, right = of cy,yshift=8 mm] {$Y$}; 
\node (1b) [terminal2, left = of 1y,xshift=2 mm] {1}; 
\node (1a) [terminal2, left = of 1b,xshift=-12 pt] {1}; 
\node (1) [terminal2, left = of 1a, xshift=2 mm] {1}; 
\graph{
    (ui1.east) ->[vh path]       (1);
      (1) -> (1a);
  (1a) ->[dashed]  (1b);
   (1b) -> (1y);
   };
\node (3y) [font=\scriptsize, blue, right = of cy,yshift=-4 mm] {$Y$}; 
\node (3b) [terminal2, left = of 3y,xshift=2 mm] {1}; 
\node (3a) [terminal2, left = of 3b,xshift=-12 pt] {0}; 
\node (3) [terminal2, left = of 3a, xshift=2 mm] {0}; 
\graph{
    (ui1.east) ->[vh path]       (3);
      (3) -> (3a);
  (3a) ->[dashed]  (3b);
   (3b) -> (3y);
   };
\node (db) [font=\scriptsize, below = of 1b, yshift=10pt] {$\vdots$}; 
\node (da) [font=\scriptsize, below = of 1a, yshift=10pt] {$\vdots$}; 
\node (d) [font=\scriptsize, below = of 1, yshift=10pt] {$\vdots$};   
\node (dy) [font=\scriptsize, right = of db,  xshift=-4.5pt] {$\vdots$}; 
\end{scope}
\begin{scope}[yshift=-65pt,xshift=200pt]
\node(cy){};
\node (ui1) [rectangle, top color=white, bottom color=black!20, draw, very thick, draw=black, left= of cy, xshift=-24pt] {\kern 10 pt};
\node (0y) [font=\scriptsize, blue, right = of cy,yshift=-8 mm] {$Y$}; 
\node (0b) [terminal2, left = of 0y,xshift=2 mm] {0}; 
\node (0a) [terminal2, left = of 0b,xshift=4.5 pt] {0}; 
\node (0) [terminal2, left = of 0a, xshift=2 mm] {0}; 
\node (labela) [below = of 0a,yshift=-3pt,xshift=10pt] {(b)};     \graph{
 (ui1.east) ->[vh path]       (0);
  (0) -> (0a);
  (0a) ->[dashed]  (0b);
   (0b) -> (0y);
};
\node (2y) [font=\scriptsize, blue, right = of cy,yshift=4 mm] {$Y$}; 
\node (2b) [terminal2, left = of 2y,xshift=2 mm] {0}; 
\node (2a) [terminal2, left = of 2b,xshift=4.5  pt] {1}; 
\node (2) [terminal2, left = of 2a, xshift=2 mm] {1}; 
\graph{
    (ui1.east) ->[vh path]       (2);
      (2) -> (2a);
  (2a) ->[dashed]  (2b);
   (2b) -> (2y);
   };
\node (1y) [font=\scriptsize, blue, right = of cy,yshift=8 mm] {$Y$}; 
\node (1b) [terminal2, left = of 1y,xshift=2 mm] {1}; 
\node (1a) [terminal2, left = of 1b,xshift=4.5  pt] {1}; 
\node (1) [terminal2, left = of 1a, xshift=2 mm] {1}; 
\graph{
    (ui1.east) ->[vh path]       (1);
      (1) -> (1a);
  (1a) ->[dashed]  (1b);
   (1b) -> (1y);
   };
\node (3y) [font=\scriptsize, blue, right = of cy,yshift=-4 mm] {$Y$}; 
\node (3b) [terminal2, left = of 3y,xshift=2 mm] {1}; 
\node (3a) [terminal2, left = of 3b,xshift=4.5 pt] {0}; 
\node (3) [terminal2, left = of 3a, xshift=2 mm] {0}; 
\graph{
    (ui1.east) ->[vh path]       (3);
      (3) -> (3a);
  (3a) ->[dashed]  (3b);
   (3b) -> (3y);
   };
\node (db) [font=\scriptsize, below = of 1b, yshift=10pt] {$\vdots$}; 
\node (da) [font=\scriptsize, below = of 1a, yshift=10pt] {$\vdots$}; 
\node (d) [font=\scriptsize, below = of 1, yshift=10pt] {$\vdots$}; 
\node (dy) [font=\scriptsize, right = of db,  xshift=-4.5pt] {$\vdots$}; 
\node (dd1) [xshift=-20,yshift=-33] {$\ddots$};  \node (dd2) [xshift=16,yshift=-33] {$\ddots$};   \end{scope}
\begin{scope}[yshift=-5pt]
\foreach \x in {0,16}
{
\begin{scope}[yshift=-3.76*\x pt, xshift=\x pt]
\node(cy){};
\node (ui1) [rectangle, top color=white, bottom color=black!20, draw, very thick, draw=black, left= of cy, xshift=-40pt] {\kern 10 pt};
\node (0y) [inner sep=1.5pt, font=\scriptsize, blue, right = of cy,yshift=-8 mm] {$Y$}; 
\node (0b) [terminal2, left = of 0y,xshift=2 mm] {0}; 
\node (0a) [terminal2, left = of 0b,xshift=-12 pt] {0}; 
\node (0) [terminal2, left = of 0a, xshift=2 mm] {0}; 
\graph{
 (ui1.east) ->[vh path]       (0);
  (0) -> (0a);
  (0a) ->[dashed]  (0b);
   (0b) -> (0y);
};
\node (2y) [inner sep=1.5pt, font=\scriptsize, blue, right = of cy,yshift=4 mm] {$Y$}; 
\node (2b) [terminal2, left = of 2y,xshift=2 mm] {0}; 
\node (2a) [terminal2, left = of 2b,xshift=-12 pt] {1}; 
\node (2) [terminal2, left = of 2a, xshift=2 mm] {1}; 
\graph{
    (ui1.east) ->[vh path]       (2);
      (2) -> (2a);
  (2a) ->[dashed]  (2b);
   (2b) -> (2y);
   };
\node (1y) [inner sep=1.5pt, font=\scriptsize, blue, right = of cy,yshift=8 mm] {$Y$}; 
\node (1b) [terminal2, left = of 1y,xshift=2 mm] {1}; 
\node (1a) [terminal2, left = of 1b,xshift=-12 pt] {1}; 
\node (1) [terminal2, left = of 1a, xshift=2 mm] {1}; 
\graph{
    (ui1.east) ->[vh path]       (1);
      (1) -> (1a);
  (1a) ->[dashed]  (1b);
   (1b) -> (1y);
   };
\node (3y) [inner sep=1.5pt, font=\scriptsize, blue, right = of cy,yshift=-4 mm] {$Y$}; 
\node (3b) [terminal2, left = of 3y,xshift=2 mm] {1}; 
\node (3a) [terminal2, left = of 3b,xshift=-12 pt] {0}; 
\node (3) [terminal2, left = of 3a, xshift=2 mm] {0}; 
\graph{
    (ui1.east) ->[vh path]       (3);
      (3) -> (3a);
  (3a) ->[dashed]  (3b);
   (3b) -> (3y);
   };
\node (db) [font=\scriptsize, below = of 1b, yshift=10pt] {$\vdots$}; 
\node (da) [font=\scriptsize, below = of 1a, yshift=10pt] {$\vdots$}; 
\node (d) [font=\scriptsize, below = of 1, yshift=10pt] {$\vdots$}; 
\node (dy) [font=\scriptsize, right = of db,  xshift=-7pt] {$\vdots$}; 
\end{scope}
}
\node (dd1) [xshift=-10,yshift=-92] {$\ddots$};
\node (dd2) [xshift=30,yshift=-92] {$\ddots$};
\node (labela) [below = of 0a,yshift=-3pt,xshift=10pt] {(a)};     \end{scope}
\end{tikzpicture}
\caption{(a) A {\em history-restricted} marginal structural model consists of multiple marginal structural models for different endpoints, conditioned on a history of a fixed length.
(b) A {\em history-adjusted} marginal structural model consists of multiple marginal structural models at different times for the same endpoint. History-adjusted models are over-parameterized and thus potentially may imply multiple contradictory estimates for the same counterfactual mean.\label{fig:ha}}
\end{figure}

This is an interesting paper on an important problem. As internet-enabled
devices become increasingly ubiquitous, manufacturers and developers are
employing randomized experiments to optimize the performance of their
products. The methods presented have close relationships to others 
in the literature, in particular to a series of papers by Robins, Hern{\'{a}}n and collaborators on analyzing observational studies as a series of
randomized trials \citep{2005robins:womens-health,hernan2008observational,hernan-robins-nejm-2017},
also described as \emph{emulating} a desired target randomized trial 
\citep{robins:2016:bigdata:aje}. There is also a close relationship to the history-restricted marginal structural models (MSM) of \citet{neugebauer2007} and the history-adjusted MSM of \citet{JoffeM2001PMSM,vanderlaan:petersen:joffe:2005:ha-msm,petersen:2007:ha-msm}. See Figures \ref{fig:msm-snmm}--\ref{fig:ha} for graphical depictions of these models.
However, there are
important differences between the context in which all of the above models were
proposed and that considered by Qian et al.; these differences have
methodological implications.

To the best of our understanding a causal contrast is an excursion effect
according to Qian et al.'s conception if it is:

\begin{itemize}
\item[(I)] a contrast between the distributions of the potential outcomes
under two \textquotedblleft time-varying treatments [regimes] occurring over
an interval of time extending into the future,\textquotedblright\ that
deviate from the treatment protocol;

\item[(II)] a contrast that is \textquotedblleft marginal over prior
treatment assignments\textquotedblright .
\end{itemize}

As we show in \S \ref{sec:obsasrt} below, analyses of contrasts with both of
these characteristics were also considered in the above papers 
  by Hern\'{a}n et al. 
In addition, as noted by the authors and further explored in \S \ref{sec:hrmsm} below, a
similar marginalization idea to (II) was proposed
in the literature on history-adjusted and restricted marginal structural models.

\section{Relation to `Observational Studies Analyzed as Randomized Trials'}

\label{sec:obsasrt}

Although widely applied in the epidemiologic and medical literature, the
analytic methods in the above papers of Hern\'{a}n and Robins are less known
to the statistical literature than alternative methods for analyzing causal
effects of time-varying treatments such as doubly robust g-estimation of
structural nested models and inverse probability of treatment weighting and
doubly-robust estimation of marginal structural models including the history
restricted and adjusted versions. It is our hope that, by demonstrating the close 
correspondence between Qian's methodology for the analysis of sequential
randomized experiments and Hern\'{a}n and Robins's methodology for analyzing
observational studies, this commentary will serve to enhance the
understanding of their commonalities and stimulate further methodological
research. To demonstrate this correspondence, we begin by reviewing the formal
counterfactual framework for studying the causal effects of time-varying
treatments \citep{robins86new}. We will largely follow the development of \citet{robins2009longitudinal}.

A sequentially randomized experiment (SRE) is a randomized experiment in
which the treatment $A_{t}$ at each successive times $t$ is randomly
assigned with known randomization probabilities $p_{t}(H_{t})$ that, by
design, may depend on a subject's past treatment and covariate history
$H_{t}=\left( \overline{A}_{t-1},\overline{X}_{t}\right) $ up to time $t$; such
trials were referred to as \textit{alternative designed RCTs} in \citep{robins86new}. The micro-randomized trial of \citet{qian2019estimating}
are thus SREs. Following Qian \textit{et al.}, in a slight departure from
the ordinary meaning of the protocol of a trial, we refer to the set of
treatment probabilities $\left\{ p_{t}(H_{t});t=1,\ldots ,T\right\} $ as the 
\emph{protocol} of the SRE.

The identifying assumptions 1--3 of \citet{qian2019estimating}, namely
consistency, positivity, and sequential ignorability, will quite generally
hold in a SRE. A key insight in \cite{robins86new} was to recognize that the
three identifying assumptions could hold in an observational study and when
they did so, the observational study can be conceptualized as a sequentially
randomized experiment (run by nature), except that the protocol
probabilities $p_{t}(H_{t})$ are unknown and therefore must be estimated
from the data. However, in an observational study the assumption of
sequential ignorability is not guaranteed by design and is not subject to
empirical verification. The best one can do is to use subject-matter
knowledge in the hope of collecting data in $X_{t}$ on sufficiently many
potential time-dependent confounders to plausibly satisfy the identifying
assumptions 1--3.

A \emph{deterministic} treatment regime is a set of functions (rules) $g=\{g_{1}({x}_{1}),\ldots ,g_{T}(\overline{x}_{T},\overline{a}_{T-1})\}$
which specify treatment $a_{t}$ at time $t$ as a deterministic function $g_{t}$ of the subject's past data $h_{t}=(\overline{x}_{t}, \overline{a}_{t-1})$. A \emph{random} regime replaces the functions $g_{t}$ by
conditional densities specifying the distribution of $A_{t}$ given $\overline{X}_{t},\overline{A}_{t-1}$ under the regime. We call a regime 
\emph{dynamic} if either $g_{t}$ or the corresponding conditional
distribution depends on $\overline{X}_{t}$, and \emph{non-dynamic} or \emph{static} 
 otherwise. Using this terminology an SRE is a dynamic random regime. We
denote the potential outcomes under a regime $g$ as $O(g)$. We note that in
a medical context the optimal treatment strategy must be a dynamic regime
whenever a drug treatment, such as a chemo-therapeutic agent, has serious
associated toxicities; whenever a patient develops a severe toxicity such as a low white cell
count, it is essential to temporarily discontinue the drug.

It follows from the above that a contrast between the distributions of $O(g)$
and $O(g^{\prime })$ under regimes $g$ and $g^{\prime }$ thus trivially
corresponds to (I) in our understanding of an excursion effect. We now turn
our attention to the estimation of excursion effects marginalized over prior
treatment assignments (II). We first review methods that use observational
data to emulate a series of hypothetical randomized target trials as
introduced in the aforementioned papers of Hern\'{a}n and Robins. 
A novel aspect of the emulation is that each subject in the observational
data set is enrolled in all of the target trials for which she is eligible,
instead of a single trial. It is this feature that underlies the
correspondence between this methodology and that of Qian et al.

A \emph{target trial} is a RCT one would like to conduct on HMO members but
cannot due to ethical, financial and/or logistical reasons. As a specific
example, we consider emulation of target trials designed to estimate the
effect of post-menopausal hormone (PMH)\ therapy on the $\Delta $-year risk
of breast cancer in post-menopausal women who are within $10$ years of
menopause at time of randomization, are members of a large HMO, such as
Kaiser Permanente, and have not taken PMH for a year prior to enrollment.
The time index $t$ will denote years since January 1, 2000. We have
available the observational data $O=(X_{0},A_{0},\ldots
,X_{T},A_{T},X_{T+1}) $ on female HMO members contained in the HMO
electronic medical records [EMR], where $X_{0}$ includes all EMR data prior
to time $0$. We will show that it is possible to specify a target trial design such that
the causal estimand as well as the identifying formula for
and an estimator of this effect are formally identical to those described by \citet{qian2019estimating}. In order to specify the target trial design and
outcome we define the following $\left\{ 0,1\right\} $ dichotomous variables:

\begin{itemize}
\item[$A_t$:] $A_{t}=1$ if taking hormones at $t$,

\item[$D_t$:] $D_{t}=1$ if clinical breast cancer is diagnosed at or before $t$;

\item[$I_{t}^{*}$:] $I_{t}^{\ast }=0$ indicates treatment ineligibility at $t $. In our case, since PMHs are sometimes considered to be medically contraindicated in premenopausal
women or women with \ history of deep vein thrombosis (DVT) or breast
cancer, we have $I_{t}^{\ast }=0$ if DVT or breast cancer has occurred at or
before $t$ or if the woman is pre-menopausal;

\item[$I_{t}$:] $I_{t}=0$ indicates the subject is ineligible for a target
trial with enrollment at $t$; in our case $I_{t}=0$ if and only if at least one of
the following is true: the patient is treatment ineligible ($I_{t}^{\ast
}=0) $, the women is greater than 10 years from menopause, or the patient
has been on PMH during the past year so that $A_{t-1}=1$. \end{itemize}

We begin by considering a single target trial in which trial eligible HMO
members are enrolled and randomized on a specific calendar date $t$ years
from 1 January 2000. For the sake of concreteness we take $t=4$. Consider a
woman who is trial eligible at $t$ so that $I_{t}=1$. 
The trial outcome $Y_{t,\Delta }$ is development of clinical breast cancer
within $\Delta $ years from randomization i.e. $Y_{t,\Delta }=D_{t+\Delta }$. She is randomized with probability $1/2$ to the arm $G=g^{\ast }$ or $G=g^{\prime }$, where $g^{\ast}$ and $g^{\prime }$ are the treatment regimes being
compared in the target trial. As an example, since women are often 
prescribed PMH for one year, two natural regimes to compare would be $g^{\ast }=(\overline{A}_{t-1},1,\overline{0}_{\Delta -1})$ corresponding to
one year of PMH followed by $\Delta\!-\!1$ years without, and $g^{\prime }=(\overline{A}_{t-1},0,\overline{0}_{\Delta -1})$, corresponding to no PMH for
the next $\Delta $ years.

\begin{figure}
\begin{tikzpicture}[every new ->/.style={shorten >=1pt},
graphs/every graph/.style={edges=rounded corners},
vh path/.style={to path={-- ++(3.5pt,0) |- (\tikztotarget)}},
vh path2/.style={to path={|- (\tikztotarget)}}
]
\begin{scope}
\node (brace) [font=\scriptsize,yshift=20pt,xshift=-58pt] {$\overbrace{\hbox{\kern 80pt}}^{\textstyle \Delta}$}; 
\foreach \x in {0,6,12}
{
\FPeval{Width}{(10)};
\begin{scope}[yshift=-5.5*\x pt, xshift=4*\x pt]
\node(cy){};
\node (0y) [font=\scriptsize, blue, above = of cy,yshift=-6mm] {$Y$}; 
\node (0b) [terminal, left = of 0y,xshift=3 pt] {0}; 
\node (0a) [terminal, left = of 0b,xshift=-16 pt] {0}; 
\node (0) [terminal, green!50!black, left = of 0a, xshift=4pt] {1}; 
\node (1y) [font=\scriptsize, blue, below = of cy,yshift=6mm] {$Y$}; 
\node (1b) [terminal, left = of 1y,xshift=3 pt] {0}; 
\node (1a) [terminal, left = of 1b,xshift=-16 pt] {0}; 
\node (1) [terminal, red, left = of 1a,xshift=4pt] {0}; 
\node (ui1) [rectangle, top color=white, bottom color=black!20, draw, very thick, draw=black, left= of cy, xshift=-92 pt] {\kern \Width pt};
\graph{
 (ui1.east) ->[vh path,green!50!black]       (0);
  (ui1.east) ->[vh path,red]       (1);
   (0) ->  (0a);
  (0a) ->[dashed]  (0b);
   (0b) -> (0y);
  (1) ->  (1a);
  (1a) ->[dashed]  (1b);
   (1b) -> (1y);
  };
\end{scope} 
}
\node (dd1) [xshift=-36,yshift=-84] {$\ddots$};
\node (dd1) [xshift=32,yshift=-84] {$\ddots$};
\node (labela) [xshift=-40.5,yshift=-120]{(a)};
\end{scope}
\begin{scope}[xshift=275pt]
\FPeval{Width}{(8)};
\begin{scope}[yshift=50 pt, xshift=-120pt]
\node(cy){};
\node (0y) [inner sep=1pt, font=\scriptsize, blue, above = of cy,yshift=-6mm] {$Y$}; 
\node (0b) [terminal2, green!50!black,  left = of 0y,xshift=5pt] {1}; 
\node (1y) [inner sep=1pt, font=\scriptsize, blue, below = of cy,yshift=6mm] {$Y$}; 
\node (1b) [terminal2, red, left = of 1y,xshift=5pt] {0}; 
\node (ui1) [rectangle, top color=white, bottom color=black!20, draw, very thick, draw=black, left= of cy, xshift=-12 pt] {\kern \Width pt};
\graph{
 (ui1.east) ->[vh path2, green!50!black]       (0b);
  (ui1.east) ->[vh path2, red]       (1b);
   (0b) -> (0y);
   (1b) -> (1y);
};
\end{scope}
\begin{scope}[yshift=30 pt,xshift=-103pt]
\node(cy){};
\node (0y) [inner sep=1pt, font=\scriptsize, blue, above = of cy,yshift=-6mm] {$Y$}; 
\node (0b) [terminal2, green!50!black, left = of 0y,xshift=5pt] {1}; 
\node (1y) [inner sep=1pt, font=\scriptsize, blue, below = of cy,yshift=6mm] {$Y$}; 
\node (1b) [terminal2, red, left = of 1y,xshift=5pt] {0}; 
\node (ui1) [rectangle, top color=white, bottom color=black!20, draw, very thick, draw=black, left= of cy, xshift=-12 pt] {\kern \Width pt};
\graph{
 (ui1.east) ->[vh path2, green!50!black]       (0b);
  (ui1.east) ->[vh path2, red]       (1b);
   (0b) -> (0y);
   (1b) -> (1y);
};
\end{scope}
\begin{scope}[yshift=22pt,xshift=-80pt]
\node(dd)[yshift=3pt]{$\ddots$};
\node(d1)[font=\scriptsize,xshift=-112pt,yshift=0pt]{$\Delta=1$};
\end{scope}
\begin{scope}[yshift=15pt,xshift=-10pt]
\node(cy){};
\node (0y) [inner sep=1pt, font=\scriptsize, blue, above = of cy,yshift=-6mm] {$Y$}; 
\node (0b) [terminal2, green!50!black, left = of 0y,xshift=5pt] {1}; 
\node (1y) [inner sep=1pt, font=\scriptsize, blue, below = of cy,yshift=6mm] {$Y$}; 
\node (1b) [terminal2, red,  left = of 1y,xshift=5pt] {0}; 
\node (ui1) [rectangle, top color=white, bottom color=black!20, draw, very thick, draw=black, left= of cy, xshift=-12 pt] {\kern \Width pt};
\graph{
 (ui1.east) ->[vh path2, green!50!black]       (0b);
  (ui1.east) ->[vh path2, red]       (1b);
   (0b) -> (0y);
   (1b) -> (1y);
};
\end{scope}
\begin{scope}[yshift=-5pt,xshift=7pt]
\node(cy){};
\node (0y) [inner sep=1pt, font=\scriptsize, blue, above = of cy,yshift=-6mm] {$Y$}; 
\node (0b) [terminal2, green!50!black, left = of 0y,xshift=5pt] {1}; 
\node (1y) [inner sep=1pt, font=\scriptsize, blue, below = of cy,yshift=6mm] {$Y$}; 
\node (1b) [terminal2, red, left = of 1y,xshift=5pt] {0}; 
\node (ui1) [rectangle, top color=white, bottom color=black!20, draw, very thick, draw=black, left= of cy, xshift=-12 pt] {\kern \Width pt};
\graph{
 (ui1.east) ->[vh path2, green!50!black]       (0b);
  (ui1.east) ->[vh path2, red]       (1b);
   (0b) -> (0y);
   (1b) -> (1y);
};
\end{scope}
\begin{scope}[yshift=-15pt,xshift=-180 pt]
\node(left){};
\node(right) [xshift=200pt]{};
\graph{
(left) -- (right)
};
\end{scope}
\begin{scope}[yshift=-25 pt, xshift=-120+17 pt]
\node(cy){};
\node (0y) [inner sep=1pt, font=\scriptsize, blue, above = of cy,yshift=-6mm] {$Y$}; 
\node (0b) [terminal2, red, green!50!black, left = of 0y,xshift=5pt] {1}; 
\node (0a) [terminal2, red, green!50!black, left = of 0b,xshift=5pt] {1}; 
\node (1y) [inner sep=1pt, font=\scriptsize, blue, below = of cy,yshift=6mm] {$Y$}; 
\node (1b) [terminal2, red, left = of 1y,xshift=5pt] {0}; 
\node (1a) [terminal2, red, left = of 1b,xshift=5pt] {0}; 
\node (ui1) [rectangle, top color=white, bottom color=black!20, draw, very thick, draw=black, left= of cy, xshift=-12-17 pt] {\kern \Width pt};
\graph{
 (ui1.east) ->[vh path2, green!50!black]       (0a);
  (ui1.east) ->[vh path2, red]       (1a);
   (0a) ->[green!50!black] (0b);
   (1a) ->[red] (1b);
   (0b) -> (0y);
   (1b) -> (1y);
};
\end{scope}
\begin{scope}[yshift=-45 pt, xshift=-120+17+17 pt]
\node(cy){};
\node (0y) [inner sep=1pt, font=\scriptsize, blue, above = of cy,yshift=-6mm] {$Y$}; 
\node (0b) [terminal2, green!50!black, left = of 0y,xshift=5pt] {1}; 
\node (0a) [terminal2, green!50!black, left = of 0b,xshift=5pt] {1}; 
\node (1y) [inner sep=1pt, font=\scriptsize, blue, below = of cy,yshift=6mm] {$Y$}; 
\node (1b) [terminal2, red, left = of 1y,xshift=5pt] {0}; 
\node (1a) [terminal2, red, left = of 1b,xshift=5pt] {0}; 
\node (ui1) [rectangle, top color=white, bottom color=black!20, draw, very thick, draw=black, left= of cy, xshift=-12-17 pt] {\kern \Width pt};
\graph{
 (ui1.east) ->[vh path2, green!50!black]       (0a);
  (ui1.east) ->[vh path2, red]       (1a);
   (0a) ->[green!50!black] (0b);
   (1a) ->[red] (1b);
   (0b) -> (0y);
   (1b) -> (1y);
};
\end{scope}
\begin{scope}[yshift=-58pt,xshift=-72pt]
\node(dd)[yshift=8pt]{$\ddots$};
\node(d2)[font=\scriptsize,xshift=-120pt,yshift=10pt]{$\Delta=2$};
\end{scope}
\begin{scope}[yshift=-65 pt, xshift=7 pt]
\node(cy){};
\node (0y) [inner sep=1pt, font=\scriptsize, blue, above = of cy,yshift=-6mm] {$Y$}; 
\node (0b) [terminal2, green!50!black, left = of 0y,xshift=5pt] {1}; 
\node (0a) [terminal2, green!50!black, left = of 0b,xshift=5pt] {1}; 
\node (1y) [inner sep=1pt, font=\scriptsize, blue, below = of cy,yshift=6mm] {$Y$}; 
\node (1b) [terminal2, red, left = of 1y,xshift=5pt] {0}; 
\node (1a) [terminal2, red, left = of 1b,xshift=5pt] {0}; 
\node (ui1) [rectangle, top color=white, bottom color=black!20, draw, very thick, draw=black, left= of cy, xshift=-12-17 pt] {\kern \Width pt};
\graph{
 (ui1.east) ->[vh path2, green!50!black]       (0a);
  (ui1.east) ->[vh path2, red]       (1a);
   (0a) ->[green!50!black] (0b);
   (1a) ->[red] (1b);
   (0b) -> (0y);
   (1b) -> (1y);
};
\end{scope}
\begin{scope}[yshift=-76pt,xshift=-180 pt]
\node(left){};
\node(right) [xshift=200pt]{};
\graph{
(left) -- (right)
};
\end{scope}
\begin{scope}[yshift=-81pt,xshift=-72pt]
\node(dd){$\vdots$};
\end{scope}
\begin{scope}[yshift=-93pt,xshift=-180 pt]
\node(left){};
\node(right) [xshift=200pt]{};
\graph{
(left) -- (right)
};
\end{scope}
\begin{scope}[yshift=-104 pt, xshift=7 pt]
\node(dt)[font=\scriptsize,xshift=-198pt,yshift=0pt]{$\Delta=T$};
\node(cy){};
\node (0y) [inner sep=1pt, font=\scriptsize, blue, above = of cy,yshift=-6mm] {$Y$}; 
\node (0b) [terminal2, green!50!black, left = of 0y,xshift=5pt] {1}; 
\node (0a) [terminal2, green!50!black, left = of 0b,xshift=-88pt] {1}; 
\node (0) [terminal2, green!50!black, left = of 0a,xshift=5pt] {1}; 
\node (1y) [inner sep=1pt, font=\scriptsize, blue, below = of cy,yshift=6mm] {$Y$}; 
\node (1b) [terminal2, red, left = of 1y,xshift=5pt] {0}; 
\node (1a) [terminal2, red, left = of 1b,xshift=-88pt] {0}; 
\node (1) [terminal2, red, left = of 1a,xshift=5pt] {0}; 
\node (ui1) [rectangle, top color=white, bottom color=black!20, draw, very thick, draw=black, left= of cy, xshift=-12-17-110 pt] {\kern \Width pt};
\graph{
 (ui1.east) ->[vh path2, green!50!black]       (0);
  (ui1.east) ->[vh path2, red]       (1);
   (0) ->[green!50!black] (0a);
   (1) ->[red] (1a);
   (0a) ->[dashed, green!50!black] (0b);
   (1a) ->[dashed, red] (1b);
   (0b) -> (0y);
   (1b) -> (1y);
};
\end{scope}
\node (labela) [xshift=0,yshift=-5 pt, below= of dd]{(b)};
\end{scope}
\end{tikzpicture}
\caption{(a) The Qian et al. excursion model consisting of contrasts for a final blip of treatment with a fixed time $\Delta$ to outcome $Y=Y_{t,\Delta}$; (b) The analysis of \citet{2005robins:womens-health,hernan2008observational} estimates the full survival curve and hence models all possible trial durations $\Delta$ to the outcome $Y=Y_{k}$; see Eq.~(\ref{eq:g-blip}); chosen estimands were contrasts between always receiving treatment versus never receiving treatment. Both models are conditioned on a history of a fixed length.\label{fig:qian-hernan}}
\end{figure}
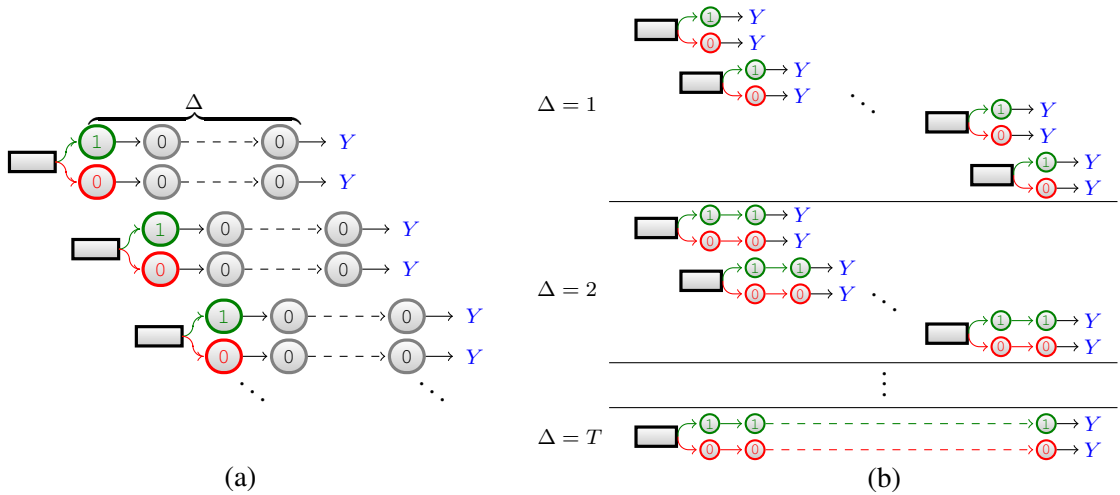

We take as our contrast the $t$-specific counterfactual blip
function between the above regimes $g^{\ast }$ and $g^{\prime }$ on the multiplicative scale: 
\begin{equation}
\beta _{t,\Delta }(S_{t})=\log \frac{\E\{Y_{t,\Delta }(\overline{A}_{t-1},1,\overline{0}_{\Delta -1}\,)\,|\,S_{t}(\overline{A}_{t-1}),I_{t}(\overline{A}_{t-1})\!=\!1\}}{\E\{Y_{t,\Delta }(\overline{A}_{t-1},0,\overline{0}_{\Delta
-1})\,|\,S_{t}(\overline{A}_{t-1}),I_{t}(\overline{A}_{t-1})\!=\!1\}}.
\label{eq:excurs-contrast}
\end{equation}Here $S_{t}(\overline{A}_{t-1})=S_{t}\subset H_{t}$ is a vector of
covariates chosen by an investigator wishing to determine whether these
covariates modify the effect of treatment on this scale. Note that the RHS
of (\ref{eq:excurs-contrast}) was written as $\beta _{M}\{t,S_{t}(\overline{A}_{t-1})\}$ by \citet{qian2019estimating}; we write $t$ and $\Delta $ as
subscripts because, to this point, we are considering $t$ and $\Delta $
fixed; see Figure \ref{fig:qian-hernan}(a).

Contrast (\ref{eq:excurs-contrast}) is an excursion effect in both sense (I)
and (II) since it does not condition on all of $H_{t}$. Had we actually
conducted this target trial, the contrast (\ref{eq:excurs-contrast}) would
then be identified from the target trial data $(O,G)$ by 
\begin{equation*}
\log \frac{\E\{Y_{t,\Delta }|\,S_{t},I_{t}=1,G=g^{\ast }\}}{\E\{Y_{t,\Delta
}\,|\,S_{t},I_{t}=1,G=g^{\prime }\}}.
\end{equation*}

However, by definition, the variable $G$ does not exist in the observational
data $O$ since there was no randomization at $t=4$, or indeed, at any other
time! Hence there is no particular reason to privilege $t=4$ rather than any
other value of $t$. That is, for the particular choice of regimes $g$ and $g^{\prime }$ above, the observational data can be used to emulate a \emph{series} of $T\!-\!\Delta\!+\!2$ target trials with enrollment at $t=0,\ldots
,T\!-\!\Delta\!+\!1$ and estimand $\beta _{\Delta }(t,S_{t})$, where $\Delta$
remains fixed. Each woman in the observational data is enrolled in each of $T\!-\!\Delta\!+\!2$ targets trials for which she satisfies the eligibility criteria $I_{t}=1$.

Under the identifying assumptions 1--3 of \citet{qian2019estimating}, the
parameters $\beta _{\Delta }(t,S_{t})$ are identified from the observational
data $O$. The identifying formula is formally the same as that given in Eq.~(4) of
Qian et al. It follows that if we imposed the parametric model of Qian et
al. for $\beta _{\Delta }(t,S_{t})$ given by their Eq.~(9) indexed by $\beta$
and also their nuisance model indexed by $\alpha$ then we could use the
estimating function given by their Eq. (10), except, because we are in an
observational study, we must estimate the unknown treatment probabilities $p_{j}( h_{j})$ from the data. If our estimates of $p_{t}( h_{t})$, $t=0,\ldots ,T$, are consistent then the estimator of $(\beta,\alpha)$ given
by Qian et al. Eq.~(10) will be consistent.

However, because consistency of our estimators of $p_{t}(h_{t})$ cannot be
assured, we would like to use a doubly robust estimator of $\beta _{\Delta
}(t,S_{t})$. The estimator of Qian et al. Eq.~(10) is not doubly robust. 
This is due to the fact that in the final product of the expression in
Eq.~(11) for the weight $J_{t}$, the projection of the terms $\mathbbm{1}(A_{j}=0)/\{1-p_{j}(H_{j})\}$ from $t+1$ to $t+\Delta -1$ onto the scores
for treatment have not been subtracted off. Even when, as in their case, the 
$p_{j}(H_{j})$ are known, subtracting off this projection would generally
increase efficiency; see, for example, \citet{robins:rotnitzky:recovery:1992,
pmid20019887}.

Qian et al.~only considered the blip to zero contrasts (\ref {eq:excurs-contrast}) between the counterfactual outcome $Y_{t,\Delta }(\overline{A}_{t-1},1,\overline{0}_{\Delta -1}\,)$ under the static regime $g^*=(\overline{A}_{t-1},1,\overline{0}_{\Delta -1})$ and the outcome $Y_{t,\Delta }(\overline{A}_{t-1},0,\overline{0}_{\Delta -1})$ under the
static regime $g^{\prime }=(\overline{A}_{t-1},0,\overline{0}_{\Delta -1})$,
although they also note that their results can be extended to contrasts
between other (identified) excursions.

To the best of our understanding, for Qian et al. the variables $I_t$ and $I_t^*$ are identical and therefore treatment is withheld when $I_t=0$. In that case,
as implicitly recognized by Qian et al., the two regimes occurring in (\ref {eq:excurs-contrast}) are the only static regimes that are identified
without further assumptions.
This is because any other static regime $\tilde{g}$ will have $a_m=1$ for some $m>t$. However, if $I_m^*=I_m=0$ with
positive probability under $\tilde{g}$ then the counterfactual outcome will
not be identified since $I_m=0$ deterministically implies $A_m=0$ and thus
positivity fails. Note that the blip excursion (\ref{eq:excurs-contrast}) is
only identifiable without further assumptions \emph{because} there is
``one-sided compliance'', so that if $I^*_{m}=0$ for $m>t$, then they receive
treatment $A_{m}=0$.  For further
discussion of this point in a medical setting, see \citet{hernan-robins-nejm-2017}.

\section{Relation to Varieties of Marginal Structural Models}

\label{sec:hrmsm}

\subsection{History-Restricted Marginal Structural Models}

As noted by Qian et al. the problem context is similar to that for which the
history restricted marginal structural models (HR-MSMs) \citep{neugebauer2007} were developed. Here we show that, as Qian et al.
suggest, these models can be viewed as identifying a large number of
excursion effects. To avoid complexity (notational and otherwise) that
obscures the central point we wish to make in this section, we shall assume
that $I_{m}^{\ast }=I_{m}$=$1$ with probability $1$ so that we can restrict the discussion to
static regimes. A HR-MSM is a model for $\E\{Y_{t,\Delta }(\overline{A}_{t-1},a_{t},\ldots ,a_{t+\Delta })\,|\,S_{t}(\overline{A}_{t-1}),I_{t}(\overline{A}_{t-1})\!=\!1\}$ all $t\in \left\{ 1,\ldots ,T\!-\!\Delta\! +\!1\right\} 
$, all $(a_{0},\ldots ,a_{T})\in \{0,1\}^{T+1}\ $and a single pre-specified $\Delta$; see Figure \ref{fig:ha}(a).

To see the connection with the model of Qian et al. consider a simple HR-MSM
that is linear in cumulative exposure on a log scale with parameters $(\alpha_t,\beta_t)$: 
\begin{equation}
\log {\E\{Y_{t,\Delta }(\overline{A}_{t-1},a_{t},\ldots ,a_{t+\Delta
})\,|\,S_{t},I_{t}\!=\!1\}} =b_{t,\Delta}(S_{t};\alpha_t)+ \beta_t
\sum_{j=t}^{t+\Delta }a_{j}.  \label{eq:hr-msm}
\end{equation}The model (\ref{eq:hr-msm}) satisfies 
\begin{equation}
\beta _{t,\Delta }(S_{t}) = \log \frac{\E\{Y_{t,\Delta }(\overline{A}_{t-1},a_{t},\ldots ,a_{t+\Delta })\,|\,S_{t}(\overline{A}_{t-1}),I_{t}(\overline{A}_{t-1})\!=\!1\}}{\E\{Y_{t,\Delta }(\overline{A}_{t-1},0,
{\overline{0}}_{\Delta -1})\,|\,S_{t}(\overline{A}_{t-1}),I_{t}(\overline{A}_{t-1})\!=\!1\}}=\beta _{t}\sum_{j=t}^{t+\Delta }a_{j}.  \label{eq:hr-msm2}
\end{equation}The model (\ref{eq:hr-msm}) satisfies (II) because the contrasts (\ref {eq:hr-msm2}) are marginal over prior treatment assignments. It also
satisfies (I) in that it specifies, for \emph{every }$t$ and \emph{every}
value of $S_t$ a contrast between each of the $2^{\Delta}-1$ regimes $(a_{t},\ldots ,a_{t+\Delta })$ and $\overline{0}_{\Delta }$. As a
consequence a parametric model such as (\ref{eq:hr-msm}) is highly unlikely
to be correctly specified except under the null.

An HR-MSM, such as (\ref{eq:hr-msm}), that does not link the parameters for
different times is simply a collection of ordinary marginal structural
models that therefore can be fitted separately \citep{robins:2007:history-adjusted-msm-discussion}. Of course, they become
related if one chooses to impose stationarity assumptions, such as $\beta_t
= \beta$ for all $t$. 

\subsection{History-Adjusted Marginal Structural Models}

A \emph{History-Adjusted Marginal Structural Model} \citep{JoffeM2001PMSM,vanderlaan:petersen:joffe:2005:ha-msm,petersen:2007:ha-msm}
differs from a HR-MSM only in that, in the model definition the phrase a
``single prespecified $\Delta $'' is replaced by ``all $\Delta \in
\{1,\ldots ,T\!-\!t\!+\!1\}$,'' see Figure \ref{fig:ha}(b).

In contrast to history-restricted models, \citet{robins:2007:history-adjusted-msm-discussion} show in their appendix that in the case
where the set $S_{t}$ is the entire history $H_{t}$ then the models may be
over-parametrized and hence may be incoherent in the following sense:
a given counterfactual mean may be expressed both as a function of one
subset of the model parameters and as a different function of a second
non-overlapping subset of parameters. As shown  by Robins et al,
this implies that one could fit a mis-specified history-adjusted model and
produce two separate estimates of the mean of a particular counterfactual
regime which differ in sign, with the difference between the estimates many
standard errors from zero, hence rendering the analysis useless for
decision-making.

In fact the same phenomena may arise when we only condition 
on $S_t$. Specifically, consider a distribution satisfying $\beta_{m,\Delta}(H_m) = \beta_{m,\Delta}(S_t)$ for some $t$
and all $(m, \Delta)$ such that $m\geq t$ and $m+\Delta=k$ for some fixed $k$.
Then the argument given in the appendix of \citet{robins:2007:history-adjusted-msm-discussion} goes through unchanged. Such a distribution will always exist because the parameters $\beta_{m,\Delta}(H_m)$ are variation independent; see \S\ref{sec:multiple} below.

Prior to  \citet{robins:2007:history-adjusted-msm-discussion}, the consequential distinction between HA-MSM and HR-MSM was not recognized;  both models were referred to as HA-MSM in the literature. Robins et al. argued that the two models should be differentiated  and proposed the definitions given above, although the moniker HR-MSM was coined by \citet{neugebauer2007}. Readers should be aware that not all authors have adopted the model definitions given here.

\section{Target Trials with Multiple Endpoints}\label{sec:multiple}

In their published data analyses,
\citet{2005robins:womens-health,hernan2008observational} 
 took as the target trial a randomized controlled trial that compared the regime 
 $g^{con}=(\overline{A}_{t-1},\overline{1}_{\Delta})$ corresponding to continuous treatment  for the next 
$\Delta $ years to the regime  $g^{\prime }=(\overline{A}_{t-1},\overline{0}_{\Delta})$,
corresponding to no treatment for the next $\Delta $ years. The
corresponding contrast on the log risk ratio between these regimes scale is
thus 
\begin{equation}
\beta _{t,\Delta }^{con}(S_{t})=\log \frac{\E\{Y_{t,\Delta }(\overline{A}_{t-1},1,\overline{1}_{\Delta -1}\,)\,|\,S_{t}(\overline{A}_{t-1}),I_{t}(\overline{A}_{t-1})\!=\!1\}}{\E\{Y_{t,\Delta }(\overline{A}_{t-1},0,\overline{0}_{\Delta -1})\,|\,S_{t}(\overline{A}_{t-1}),I_{t}(\overline{A}_{t-1})\!=\!1\}}. \label{eq:beta-all}
\end{equation}
They further assumed that $I_{m}^{\ast }=1$ w.p.1 at all times $m$, so that
patients are always eligible to receive either treatment or control. Thus $\beta _{t,\Delta }^{con}(S_{t})$ is identifiable under sequential
randomization. Substantively, $Y_{t,\Delta }$ was the indicator of survival
at $t+\Delta$ and the authors wished to compare regime-specific survival curves.
Thus, as in a HA-MSM, they were interested in estimating $\beta
^{cont}(t,\Delta ,S_{t})=\beta _{t,\Delta }^{con}(S_{t})$ for all $t\in
\left\{ 1,\ldots ,T-\Delta +1\right\} $ \ and $\Delta \in \{1,\ldots ,T-t+1\}
$; see Figure \ref{fig:qian-hernan}(b).

 This raises the question of whether problems with overparametrization 
and incoherence might occur as with a HA-MSM. In fact, we can also ask this
question for the contrast $\beta (t,\Delta ,S_{t}) \equiv \beta _{t,\Delta }(S_{t})$
comparing $(\overline{A}_{t-1},1,\overline{0}_{\Delta -1})$ with 
$(\overline{A}_{t-1},0,\overline{0}_{\Delta -1})$ 
as earlier. We show that for both these contrasts incoherence does not occur.
To see this, following \citep{robins2004optimal}, we first consider the case where 
$H_{t}=S_{t}$. Then for any regime $g$, dynamic or static, we define the
regime specific blip functions:
\begin{equation}
\gamma _{t,k}^{g}({H}_{t})=\log \frac{\E\{Y_{k}(\overline{A}_{t-1},a_{t}=1,\underline{g}_{t+1}\,)\,|\,{H}_{t}\}}{\E\{Y_{k}(\overline{A}_{t-1},a_{t}=0,\underline{g}_{t+1})\,|\,{H}_{t}\}},\quad \hbox{for }t=1,\ldots ,k;  \label{eq:g-blip}
\end{equation}where we have reparametrized $\left\{ t,\Delta \right\} $ as $\left(
t,k\right) $ with $k=t+\Delta$, $k\in \left\{ 2,\ldots,T\right\}$; the potential outcome $Y_k
(\cdot, \cdot ,\underline{g}_{t+1})$ indicates that regime $g$ is
followed from $t+1$ onwards.
Further, if $\gamma _{t,k}({H}_{t})=0$
with probability 1 for all $t=1,\ldots ,k$, then under sequential
randomization $\E\{Y_{k}(\overline{A}_{t-1},\tilde{\underline{g}}_t)\mid {H}_{t}\}\ =\E\{Y_{k}|{H}_{t}\}$ with probability $1$ for all $t$ and identified regimes $\tilde{g}$  \citep{robins2004optimal},
hence there is no causal effect of any regime, dynamic or static.

Consider the following two special cases:\par

\begin{itemize}
\item[(1)] The dynamic regime $g_{t}( {H}_{t},\overline{A}_{t-1})
\equiv A_{t-1}$ for $t>1$.
 In this case $Y_{k}(\overline{A}_{t-1},a_{t}=1,\underline{g}_{t+1})=$ $Y_{k}(\overline{A}_{t-1},\underline{1}_{t})$,
while $Y_{k}(\overline{A}_{t-1},a_{t}=0,\underline{g}_{t+1})=$ $Y_{k}(\overline{A}_{t-1},\underline{0}_{t})$. Thus
$\gamma _{t,k}^{g}({H}_{t})$
  corresponds to (\ref{eq:beta-all}).
\item[(2)]The regime $g_{t}\left( {H}_{t},\overline{A}_{t-1}\right)
=0$. Now  $Y_{k}(\overline{A}_{t-1},a_{t}=1,\underline{g}_{t+1})=Y_{k}(\overline{A}_{t-1},a_{t}=1,\underline{0}_{t+1})$ and $Y_{k}(\overline{A}_{t-1},a_{t}=0,\underline{g}_{t+1})=Y_{k}(\overline{A}_{t-1},a_{t}=0,\underline{0}_{t+1})$.
In this case $\gamma _{t,k}^{g}({H}_{t})$ corresponds to  (\ref{eq:excurs-contrast}) considered by Qian et al. with $H_t=S_t$.
\end{itemize}

\citet{robins2004optimal} and \citet[Theorem 8.5]{robins2000sensitivity}  proved that for any regime $g$ with $H_{t}=S_{t}$, 
the set of multiplicative blip functions $\{\gamma_{t,k}^g, \hbox{ for all }t,k\}$
are variation independent provided each $Y_k$ has support on $[0,\infty)$.
The discussion of \citet{wang2017congenial} generalizes this to the case where $Y_k$ has support
on $\{0,1\}$.
Thus, when $H_{t}=S_{t}$ neither $\beta (t,\Delta
,S_{t})$ nor $\beta _{t,\Delta }^{con}(S_{t})$ can be overparametrized or
incoherent. We now argue the same is true in the general case with ${S}_{t}\subset H_{t}$. Consider the following equalities:

\begin{align*}
\exp\{\gamma _{t,k}^{g}({S}_{t})\}&\equiv \frac{\E\{Y_{k}(\overline{A}_{t-1},a_{t}=1,\underline{g}_{t+1}\,)\,|\,{S}_{t}\}}{\E\{Y_{k}(\overline{A}_{t-1},a_{t}\!=\!0,\underline{g}_{t+1}\,)\,|\,{S}_{t}\}}\\
&=\frac{\int
\E\{Y_{k}(\overline{A}_{t-1},a_{t}=1,\underline{g}_{t+1})|{H}_{t},\}{df}(
{H}_{t}| {S}_{t}) }{\int
\E\{Y_{k}(\overline{A}_{t-1},a_{t}=0,\underline{g}_{t+1}|{H}_{t})\}{df}(
{H}_{t} | {S}_{t}) } \\
&{=\frac{\int 
\exp\{ \gamma _{t,k}^{g}({H}_{t}) \}
\E\{Y_{k}(\overline{A}_{t-1},a_{t}\!=\!0,\underline{g}_{t+1})|{H}_{t} \}{df}( {H}_{t} | {S}_{t}) }{\int \E\{Y_{k}(\overline{A}_{t-1},a_{t}\!=\!0,\underline{g}_{t+1})|{H}_{t}\}{df}( {H}_{t} |{S}_{t}) }}.
\end{align*}
Hence $\exp\{\gamma _{t,k}^{g}({S}_{t})\}$
is a weighted average of  $\exp\{\gamma _{t,k}^{g}({H}_{t})\}$.
Consequently because $\exp\{\gamma _{t,k}^{g}({H}_{t})\}$
 are variation independent it follows that 
 $\exp\{\gamma _{t,k}^{g}({S}_{t})\}$
 are also variation independent and thus
coherent.

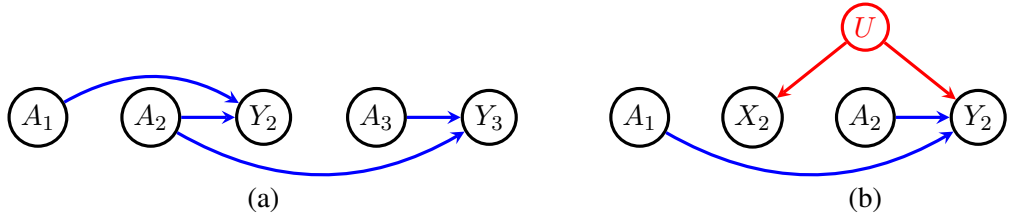
\begin{figure}[!htb]
\begin{center}
\begin{tikzpicture}
 [rv/.style={circle, draw, very thick, minimum size=5.5mm, inner sep=0.75mm}, node distance=15mm, >=stealth]
\begin{scope}
 \node[rv]  (1)              {$A_1$};
 \node[rv, right of=1] (2) {$A_2$};
 \node[rv, right of=2] (3) {$Y_2$};
 \node[rv, right of=3] (4) {$A_3$};
  \node[rv, right of=4] (5) {$Y_3$};
\path[->, color=blue,
 every node/.style={font=\sffamily\small}, style={bend right}, very thick] (2) edge node {} (5);
  \path[->, color=blue,
 every node/.style={font=\sffamily\small}, style={bend left}, very thick] (1) edge node {} (3);
  \draw[->, very thick, color=blue] (2) -- (3);
 \draw[->, very thick, color=blue] (4) -- (5);
\node[below of=3,yshift=4mm] (6) {(a)};
 \end{scope}
 \begin{scope}[xshift=8cm]
  \node[rv]  (1)              {$A_1$};
 \node[rv, right of=1] (2) {$X_2$};
 \node[rv, right of=2] (3) {$A_2$};
 \node[rv, right of=3] (4) {$Y_2$};
 \node at (3) [rv, color=red, yshift=12mm] (U) {$U$};
\path[->, color=blue,
 every node/.style={font=\sffamily\small}, style={bend right}, very thick] (1) edge node {} (4);
\draw[->, very thick, color=blue] (3) -- (4);
 \draw[<-, very thick, color=red] (2) -- (U);
 \draw[->, very thick, color=red] (U) -- (4);
  \node[below of=3,yshift=4mm] (6) {(b)};
 \end{scope}
 \end{tikzpicture}
\end{center}
\caption{(a) A causal DAG with three treatments, two outcomes and no
confounding; (b) An elaboration of the induced the DAG induced by (a) on $\{A_1,A_2,Y_2\}$, $U$ is unobserved.}
\label{fig:dag}
\end{figure}

\section{Issues arising from the excursion effect depending on the design}

The authors indicate that excursion effects should be interpreted in the
context of the existing protocol. Here we illustrate via simple examples
that changes in treatment assignment probabilities in the protocol can have
a qualitative effect on both primary and secondary analyses.

Throughout these examples we suppose the availability indicators $I_t$ are
all one. Consider the data-generating process, corresponding to the first
three nodes in the causal graph in Figure \ref{fig:dag}(a).

Note that there is no confounding between the treatments $A_1$, $A_2$, $A_3$
and the outcomes $Y_2$, $Y_3$. We made this choice to emphasize that the
above phenomena is a consequence of the interaction between the causal
effects of the treatments $A_{i-1}$ and $A_i$ on $Y_i$, for $i=2,3$.

To see this consider the following data-generating process: 
\begin{align}  \label{eq:two-step}
Y_2(a_1,a_2) \sim \hbox{Bernoulli}\{ \exp( - a_2 + 2 a_1 \cdot a_2)/4 \}.
\end{align}

Suppose treatment is assigned independently at $t=1,2$, with $\pr(A_1=1) =$ $\pr(A_2=1) = \theta$. Consider the marginal excursion effect at $t=2$, with $\Delta=1$ and $S=\emptyset$, $\beta_{t,\Delta} = \log \left[
\E\{Y_2(a_2=1)\}/\E\{Y_2(a_2=0)\}\right]$. By a simple calculation: 
\begin{align*}
\E\{Y_2(a_2=1)\} &= \sum_{a_1\in \{0,1\}} \E\{Y_2(a_1, a_2=1) \mid A_1=a_1\} \pr(A_1=a_1) \\
&= \sum_{a_1\in \{0,1\}} \E(Y_2 \mid A_1=a_1, A_2=1) \pr(A_1=a_1) \\
&= \{(1-\theta)/e + \theta e\}/4;
\end{align*}
similarly $E\{Y_2(a_2=0)\} = 1/4$. Hence: 
\begin{align*}
\beta_{t,\Delta} = \log \left[ \frac{\E\{Y_2(a_2=1)\}}{\E\{Y_2(a_2=0)\} } \right] = \log\{ (1-\theta)/e + \theta e \}.
\end{align*}
Hence $\beta_{t,\Delta}$ is negative if $\theta<1/(1+e)$, zero if $\theta=1/(1+e)$ and positive if $\theta > 1/(1+e)$.

Consequently, the meaning of the excursion effect is entirely dependent on
the \emph{prior} protocol, here the randomization probability for $A_1$,
that was in place \emph{before} the contrasted excursions commenced at $t=2$. We take it that this is the sense in which, as the authors say, excursion
effects `can be interpreted as contrasts between \emph{excursions from} the
treatment protocol' (emphasis added). In fact, this example suggests that in
certain cases, including `primary' analyses with $S_t=\emptyset$, it is 
\emph{only} possible to interpret these effects in the context of the prior
design.

Note that if instead we condition on the whole past, here $A_1$, as in a
structural nested model, we obtain the following contrast: 
\begin{align*}
\beta_{t,\Delta}(a_1) = \log \left[ \frac{\E\{Y_2(a_1,a_2=1)\}}{\E\{Y_2(a_1,a_2=0)\} } \right] = -1 + 2a_1,
\end{align*}
which is not a function of the randomization probabilities.

\bigskip

The dependence on the design also applies to secondary analyses of effect
modifiers, including those that are independent of treatment. To see this,
consider the causal graph shown in Figure \ref{fig:dag}(b), which can be
seen as an elaboration, including an additional covariate $X_2$, of the
induced sub-graph of the DAG in Figure \ref{fig:dag}(b) over $\{A_1,A_2,Y_2\} $. Further, suppose the variables are generated by the following mechanism 
\begin{align*}
Y_2(0,0),\, Y_2(1,0) &\sim_{\text{iid}} \text{Bernoulli}\{1/4\}, \\[5pt]
Y_2(0,1) \mid X_2 &\sim \text{Bernoulli}\{1/(1 + \exp(\alpha_0 - X_2)\}, \\[5pt]
Y_2(1,1) \mid X_2 &\sim \text{Bernoulli}\{1/(1 + \exp(\alpha_1 + X_2)\},
\end{align*}
where $\alpha_0 =2.666$, $\alpha_1=-0.905$ and that $X_2 \sim N(0,1)$. This
specification is such that $E\{Y_2(a_1,a_2)\}$ is still given by (\ref {eq:two-step}). For $a_1 \in \{0,1\}$ it holds that 
\begin{equation*}
\begin{split}
\E\{Y_2(a_1, 1) \mid X_2\}&= \frac{a_1}{1 + \exp(\alpha_1+ X_2)} + \frac{1-a_1}{1 + \exp(\alpha_0- X_2)}.
\end{split}\end{equation*}

\noindent Now consider the excursion effect with $S_t = X_2$ as the summary
of $H_t$: 
\begin{align}
\beta_{t,\Delta}(X_2) &= \log \frac{\E\{Y(A_1, 1) \mid X_2\}}{\E\{Y(A_1, 0)
\mid X_2\}}  \notag \\
&= \log \left\{ \theta / (1+e^{\alpha_1 + X_2}) + (1-\theta) / (1 +
e^{\alpha_0 - X_2}) \right\} + \log 4.  \label{eq:secondary-excursion}
\end{align}

We see from (\ref{eq:secondary-excursion}) that $\beta_{t,\Delta}(X_2) $ is
an increasing function of $X_2$ for $\theta$ close to $0$, 
while for $\theta$ close to $1$ it is decreasing. Consequently, in
this example, the qualitative conclusions from the secondary analysis will
also depend on the randomization probability $\theta$.

\subsection{Can excursion effects be used to modify the protocol?}

The authors say that owing to the dependence of the excursion effect on the
design this measure ``informs how the current treatment protocol might be
improved via moderation analysis on how these causal effects differ by
individual contexts.'' However, it is unclear how this would work in
practice.

Consider, for example, the marginal parameter $\beta_M$ giving the causal
effect of $A_2$ on $Y_2$ in the data generating process given by treatment (\ref{eq:two-step}). Suppose that the intention of treatment in this setting
is to reduce the occurrence of $Y=1$, so that negative values of $\beta_M$
indicate that the treatment is working as intended. Further suppose that at first, while
piloting the treatment, the experimenters use a small value of $\theta$, so $\theta<1/(1+e)$. As shown above, this will lead to a negative value of $\beta_M$. Buoyed by this news, the experimenters will likely then increase
the assignment probability so that $\theta > 1/(1+e)$. However, if they
continue to monitor $\beta_M$ they will then find that $\beta_M$ is
positive, indicating that the treatment is not working \ldots

It is also true that the excursion effects obtained from analyses of
observational studies as a series of randomized trials by Hern\'{a}n and
Robins will also depend on the `protocol', but in their setting the
`randomization probabilities' are chosen by nature and are not subject to
control by the experimenters, so the above is not an issue as there is only one design.

\section*{Acknowledgments}

This research was supported by the U.S. Office of Naval Research by grant N00014-19-1-2446.

\thispagestyle{empty} \bibliographystyle{apalike}

\end{document}